\documentclass[11pt]{article}
\hoffset = - 1.cm
\voffset = - 1.2cm
\usepackage[utf8]{inputenc}
\usepackage{amsmath}
\usepackage{amsfonts}
\usepackage{amssymb}
\usepackage{graphicx}
\usepackage{tikz}
\usepackage{tikz-cd}
\usepackage[all]{xy}
\newtheorem{theorem}{Theorem}[section]

\newtheorem{definition}[theorem]{Definition}

\newcommand{\pfrac}[2]{\frac{\partial #1}{\partial #2}}
\newcommand{\ts}{T^{*}S^{d}}
\newcommand{\tm}{T^{*} \kappa_M^{d}}

\newcommand{\hX}{\psi}
\newcommand{\hx}{\eta}
\newcommand{\hY}{\phi}
\newcommand{\hp}{\zeta}
\newcommand{\rn}{\mathbb{R}}
\newcommand{\ta}{\tilde{A}}
\newcommand{\tb}{\tilde{B}}
\newcommand{\fa}{A_4}
\newcommand{\fb}{B_4}
\newcommand{\R}[1]{\mathbb{R}^{#1}}
\newcommand{\avector}[2]{(#1_1,#1_2,\ldots,#1_{#2})}
\newcommand{\be}{\begin{equation}}
\newcommand{\ee}{\end{equation}}
\newcommand{\kdst}{$\kappa$-deformed spacetime}
\newcommand{\kd}{$\kappa$-deformed}

\newcommand{\abs}[1]{\left\lvert #1 \right\rvert}

\begin{document}

\title{Regularization of Kepler Problem in $\kappa$-spacetime}
\author{Partha Guha$^{1\,a}$, E. Harikumar$^{2\,b}$ and 
Zuhair N.S.$^{2\,c}$
\\[8pt]
$^1${\sl \small{S.N. Bose National Centre for Basic Sciences
JD Block, Sector III, Salt Lake Kolkata-700098, India}}\\[2pt]
$^2${\sl \small{School of Physics, University of Hyderabad,
Central University P O, Hyderabad-500046,
India}} \\
} 
\date{} 

\maketitle

\begin{abstract}
In this paper we regularize the Kepler problem on $\kappa$-spacetime in several different ways. First, we perform a Moser-type regularization and
then we proceed for the Ligon-Schaaf regularization to our problem. In particular, generalizing Heckman-de Laat (J. Symplectic Geom. 10, (2012), 463-473) in the 
noncommutative context we show that the Ligon-Schaaf regularization map following from an adaptation of the Moser regularization can be generalized to the 
Kepler problem on $\kappa$-spacetime.
\end{abstract}

\bigskip

\begin{quote}

\noindent
{\bf MSC primary}\,\, 53D20, 37J15, 70H05, 70H33

\bigskip



\end{quote}
{\vfill}
\footnoterule
{\noindent\small
$^{a)}${\it E-mail address:} {partha@bose.res.in} \\
$^{b)}${\it E-mail address:} {harisp@uohyd.ernet.in}  \\
$^{c)}${\it E-mail address:} {zuhairns@gmail.com}}

\section{Introduction}

It is well known that the Kepler problem refers to the bounded motion of a particle in $\mathbb{R}^{3}$ which is influenced by the 
gravitational field of a second particle fixed at the origin \cite{Cordani,Guillemin}. The Kepler problem is a completely integrable Hamiltonian system with
profound applications in physical world.
The Kepler problem has the disadvantage that it has singularities corresponding to 
collision orbits.  
Presence of singularities in a physical theory hinders a proper understanding of the corresponding system. Kepler problem, which is one 
of the most sought after problem in the history of physics, is equipped with such a singularity. Physically, the singularity point correspond 
to the point of collision between the objects under the influence of the Kepler potential. This causes serious trouble in predicting 
the dynamics of the system when there is a head-on collision. In technical terms, we say that the vector flow of the corresponding 
Hamiltonian vector field is incomplete. This issue can be overcome by regularizing the Kepler problem.

Regularization is a mathematical procedure to cure this singularity. The simplest way to carry out the regularization is to introduce a fictitious time, 
$s$ in place of physical time $t$ such that  $s$ is a function of $t$ and possibly also with a dependence on coordinates. 
The parameter $s$ so introduced is known as {\it Levi-Civita regularization parameter}\cite{levi}. A pretty clear treatment of regularizing the Kepler problem was done by Moser in his 1970 paper \cite{moser}, the
treatment of Moser relates the Kepler flow for a fixed negative energy level
to the geodesic flow on the sphere $S^n$. A lucid analysis of the geometrical aspects of Kepler problem can be found
in Milnor \cite{milnor}. In \cite{moser}, a diffeomorphism is constructed 
via stereographic projection, which establishes an equivalence between geodesic flow of Kepler field and the geodesic flow on a punctured sphere. 
This enables us to set up a one-to-one correspondence, topologically, between a negative energy surface of Kepler problem and unit tangent bundle of $S^{n}$.

An alternative approach to the regularization was found by
Ligon and Schaaf \cite{LS}. The main drawback of Moser’s regularization methods is that they handle separately each energy level and 
this disadvantage is partially removed by a regularization procedure due to Ligon and Schaaf.
The Ligon-Schaaf regularization procedure allows us to handle together all negative (resp., all positive) energy levels.  Their method is heavily dependent on computation and 
the treatment was simplified subsequently by Cushman and Bates \cite{cush1}, Cushman and Duisteermaat,\cite{cush2}, Marle \cite{Marle} and 
Hu and Santoprete \cite{HS}. In particular, Heckman and de Laat \cite{Heckman} have shown that
the Ligon-Schaaf map is the natural adaptation of the Moser map intertwining the Kepler flow
on the negative energy part of the phase space and the (geodesic)
Delaunay flow on the punctured cotangent bundle of the sphere $S^n$ in a
canonical way.

\bigskip

It is well known that, Kepler potential supplemented with Newton's theory gives a fair description of how gravitational force operate, 
at the classical level, for almost all practical purposes. But since we live in a quantum world, a complete understanding of gravity 
requires formulating a theory of gravity consistent with principles of quantum mechanics. In fact, quantum theory of gravity is a holy 
grail of theoretical physics and people have been attacking the problem of quantizing gravity along numerous paths but not yet have been 
succeeded \cite{rov,connes,asht,sw}. The reasoning that spacetime loses the operational meaning as one goes to energy scale of order of 
Planck scale have lead to introduction of spacetime with non-commuting coordinates \cite{dop}. Moyal spacetime, $\kappa$-spacetime are 
some of the well studied examples of non-commutative spacetimes \cite{connes, snyder, Riv1,Riv2}. The existence of a invariant minimum 
length have lead to the idea of doubly special relativity theories (DSR)\cite{amelino, kowalski}. The modified group transformations for 
the DSR theory is shown to satisfy an algebra known as $\kappa$-deformed algebra. Consideration of this algebra, also leads to concept 
of $\kappa$-deformed spacetime. For details see this and references therein \cite{amelino2,luk,kowalski2,ksg}.

Viewing $\kappa$-spacetime as a possible way to capture the features of spacetime at Planck length makes it interesting to study the 
geomety of $\kappa$-spacetime and hence, the force of gravity in $\kappa$-spacetime. In this paper, we would be studying the effects of 
non-commutativity on the singularity of the Kepler problem at a classical level. This would help us have a better grasp of gravity as a 
classical force, but incorporating possibility of a deformed spacetime. The $\kappa$-deformed model we study, do reduce to corect limit when 
$a\to 0$, but this model is not derived by starting from $\kappa$-deformed Riemannian geomtery.

In this paper, we will extend the regularization methods due to Moser and Ligon-Schaaf  to $\kappa$-deformed Kepler problem. A generalization of Kepler problem to the $\kappa$-deformed case is constructed \cite{zns1}, which reduces to the commutative problem as we set the deformation parameter to zero. More about studies on $\kappa$-deformed Kepler problem can be seen in \cite{hak, zns1,zns2}. The main calculation is performed via the determination of the explicit transformation formulas for stereographic projection on the $\kappa$-plane.

Organization of the paper is as follows. Section 2 outlines the basic aspects of the calculation scheme presented. In the next section we will set up a mapping from a sphere, $S^{n} \subset \R{d+1}$ onto a spatial hypersurface of $d$-dimensional $\kappa$-Minkowski spacetime, denoted as $\kappa$-$\R{d}$ expressed in terms of commutative variables. We will call such a stereographic projection as a ``semi-stereographic projection''. Once we have formulated the projection mapping, we will apply this to the regularization of Kepler problem, in section 3, using extension of Moser method to the kappa case. In section 4, we use idea of Ligon and Schaaf in the kappa-deformed situation to regularize the flow of deformed-Kepler vector field. Section 5 will contain concluding remarks and discussions about the results obtained. 
 \section{Outline \& Summary}
 In this section, we describe the calculation scheme used in this paper and also summarise the results derived. In this paper, we 
 address the issue of regularisation of Kepler problem in the $\kappa$-deformed space-time. For this, we start with the deformed Kepler 
 Hamiltonian and map it into an equivalent Hamiltonian defined entirely in terms of the commutative phase space variables
 (and deformation parameter). We achieve this by using a mapping between the coordinates of $\kappa$-deformed phase space to that 
 of commutative  phase space(see eqn.(\eqref{hatcor}) and eqn. \eqref{hatmom} below).
 
 The coordinates of the non-commutative space time do not commute with each other and these coordinates and their functions are treated as operators \cite{doug}. Another approach used is to work with functions of commutative variables, but use star-product in place of ordinary pointwise multiplication.  The star-product used in the Moyal space-time is
 \be 
f( x)*g( x) = \exp(\frac{i}{2}\theta_{mn}\pfrac{}{a_m}\pfrac{}{b_n}) f( x+ a) g(x + b)|_{a=b=0}
\ee
where the anti-symmetric tensor $\theta_{mn}$ is introduced through $[{\hat x}_m,{\hat x}_n]=i\theta_{mn}$. It is clear that this Moyal 
star product is non-local and introduces higher derivative terms. In Moyal space-time, another method used to map a problem from 
non-commutative space-time to commutative one is the Seiberg-Witten map. Many interesting results were obtained by employing
either of the above methods and usually, these results are obtained to the leading order in the non-commutative parameter $\theta$.

The Kappa-space-time is an example of a noncommutative space-time whose coordinates obey a Lie algebraic type commutative relations
(see eqn.(\eqref{kappadef}) below. Various aspects of $\kappa$-space-time and physical models on this space-time have been analysed in recent
times.  As in the case of Moyal space-time different approaches have been used to analyse $\kappa$-space-time also.  In one approach, 
differential calculus on $\kappa$-deformed space-time is used to study various effects of $\kappa$-deformation. Methods using 
star-products have also been developed and used in this context.  Another approach used is to map the coordinates of the 
$\kappa$-deformed space-time in terms of commutative ones (and deformation parameter)\cite{luk1}. This mapping has been 
extended to $\kappa$-deformed phase-space also \cite{hari}.

Here, in this paper we adopt the method of mapping coordinates of $\kappa$-deformed phase-space to that of commutative phase-space 
\cite{mal, hari}. This allows us to map $\kappa$-deformed Kepler Hamiltonian to an equivalent Hamiltonian in the commutative 
phase-space. This allows us to use well established calculation schemes. Note that the realization of the coordinates of \kd \, phase 
space given in terms of commutative phase space coordinates is parametrised by 
$\alpha, \beta, \gamma$ (see eqn. \eqref{hatcor},\eqref{hatmom}, \eqref{abc}). This realisation is derived from the Poisson bracket 
relation given in eqn.\eqref{kappadef}. As it is clear, the realisation (eqn.\eqref{hatcor},\eqref{hatmom}) is not unique and it is 
well known that choosing a realization means using a specific ordering prescription and a specific *-product. This choice also affect 
the co-algebraic structure and Drinfeld twist associated with the non-commutative theory \cite{ksg,algtwist2,algtwist3, adv.HEP}. Further we 
emphasis that in this paper we are studying the classical Kepler problem in the \kd , non-commutative space. A quantum mechanical 
generalisation will require the use of (quantum) commutation relations in place of Poisson bracket relations used in this paper 
(see eqn.\eqref{kappadef}). Use of (quantum) commutation relations will lead to further complication associated with the 
non-uniqueness of the quantization of non-canonical Poisson brackets, which will arise due to the realizations given in eqn.\eqref{hatcor}
 and \eqref{hatmom}.

Two common approaches to regularize the Kepler problem are Moser method and Ligon-Schaaf regularization. In the following sections, we incorporate these regularization techniques into the framework of \kdst. For the Moser method in the commutative spacetime, we have a transformation formula from a sphere to real plane. This transformation is then used to establish a connection between negative energy levels (bounded orbits) of Kepler problem on $\mathbb{R}^{d}$ and geodesic orbits $\mathbb{S}^{d}$. Here, in the case of \kdst , we have, instead chosen to initiate mapping between a sphere and a $\kappa$-deformed plane. This connects the phase space associated with the sphere to the phase space associated with the \kd -plane. The image function, which are non-commutative variables, obtained by this mapping are then expressed in terms of commutative variables. An alternative procedure called Ligon-Schaaf regularization, is more advantageous than the former as it connects all the negative (or positive) energy levels of Kepler problem at once to the geodesic flow on $\mathbb{S}^{d}$. Ligon-Schaaf map, acting as a canonical map, connects the phase space of Kepler problem and phase space of geodesics of sphere and relates the integrals of motion of one problem with the other. In conclusion, we have extended both the regularization techniques to \kd \, case and have demonstrated the regularization of deformed-Kepler problem under the scheme presented.

 \section{Semi-stereographic Projection}  
\section*{Notations}
\begin{itemize}
\item $(X,Y)$ denotes the phase space in $d$-dimensional Euclidean space.
\item $(u,v)$ denotes the phase space of a $d+1$-dimensional sphere.
\item $(\psi, \phi)$ or $(\eta, \zeta)$ denotes the variables in $\kappa$-Minkowski phase space.
\item $(x,p)$ or $(y,q)$ denotes the variables in Minkowski space in terms of which the $\kappa$-Minkowski phase spacetime variables are expressed as natural realization.
\end{itemize}
Before proceeding, we will briefly overview certain aspects of $\kappa$-spacetime.
\subsection{$\kappa$-spacetime}
$\kappa$-spacetime is a non-commutative spacetime with coordinates satisfying a Lie-algebraic type relation:
\begin{equation}
[\psi_0,\psi_i] = \frac{1}{\kappa} \psi_i = a \psi_i , \quad \quad [\psi_i,\psi_j] = 0. \label{kappadef}
\end{equation}
Note that the relations defined above are the Poisson brackets and should be distinguished from the commutation relations associated with quantization.
The factor $a$ is the deformation parameter of the $\kappa$-deformed spacetime. In this paper, we choose to work with non-commutative phase space variables expressed in terms of commutative variables, with the help of following realization \cite{mal,hari}.
\begin{equation}
\psi^\mu = x^{\mu}+\alpha x^{\mu}(a \cdot p)+\beta(a\cdot x)p^{\mu}+\gamma a^{\mu}(x \cdot p) \label{hatcor}.
\end{equation}
We also write the corresponding momenta as
\begin{equation}
\phi^\mu = p^{\mu}+(\alpha +\beta)(a \cdot p)p^{\mu}+\gamma a^{\mu}(p \cdot p) \label{hatmom}.
\end{equation}
Here, $\alpha, \beta$ and $\gamma$ satisfy the conditions
\begin{equation} 
\gamma-\alpha=1,~~
\alpha, \gamma, \beta \in \mathbb{R} \label{abc}
\end{equation}
These conditions are obtained by substituting eqn.\eqref{hatcor} in eqn.\eqref{kappadef} and using Poisson bracket relations among the commutative phase space variables $x^{\mu}$ and $p^{\nu}$, i.e., 
$[x^{\mu},p^{\nu}]_{P.B.} = \eta^{\mu \nu},$ $[x^{\mu},x^{\nu}]_{P.B.} =0,$ $[p^{\mu},p^{\nu}]_{P.B.} =0.$ An obvious advantage of this formalism is that we can easily work using calculus on familiar commutative space to analyze the 
consequences of non-commutativity. Another advantage being that identification of physical phenomena in commutative spacetime can be 
done in more a direct and less complicated manner. The results obtained would certainly depend on the realization used\cite{ksg,hak}. In this paper, 
we arrive at general conclusion regarding the regularization of Kepler problem on \kd \, spacetime. It has been shown that, upto first 
order in $a$, the predictions of different realizations will differ only by a numerical factor and hence the general conclusions are not 
affected \cite{hak}. 
\subsection{Stereographic projection using non-commutative coordinates}
 We would relate the d-dimensional sphere in commutative spacetime to a d-dimensional $\kappa$-Minkowski spacetime. Since we are going from a commutative spacetime to non-commutative spacetime, we will call this mapping, semi-stereographic projection. As far as calculus is concerned, $\psi$ and $\phi$ will be considered as ordinary functions and so will denote them occasionally as $\psi(x,p), \phi(x,p)$ to remind this.

 \begin{definition}
 The phase space of the unit $n$-dimensional sphere is given by
 \be 
 T^{*}S^{d} = \{ (u,v) \in \R{d+1} \bigm\vert \abs{u} =1, \, <u,v> = 0 \} 
 \ee
\end{definition}

We could now invoke the following theorem to set up a transformation  between $T^{*}S^{d}$ and $T^{*}{(\kappa_M)^{d}}$.  Here $T^{*}{(\kappa_M)^{d}}$ stands for the the cotangent bundle of $\kappa$-Minkowski spacetime expressed in terms of ordinary phase variables of the commuting space. 

\smallskip

Consider a $d$-dimensional sphere $S^{d} \subset \rn^{d+1}$ embedded in a $d+1$ dimensional space. Let $\vec{u} = \avector{u}{d+1}$ denote a generic point on the sphere with the north pole, $\vec{n} = (0, 0, \cdots, 0,1)$ taken as the reference point for the stereographic projection. The vector $\vec{X} = \avector{X}{d}$ correspond to the point to the projected space, $\rn^{d}$. $\vec{v}$ is the covector corresponding to $T^{*}S^d$ and $\vec{Y}$ be the covector of $T^{*}\mathbb{R}^d$. We chose the convention that the mapping $\vec{u} \mapsto \vec{X}$ is termed as stereographic projection, while the inverse mapping is referred to as the inverse stereographic projection. 
We will work on the generalization of the 
transformation formula for stereographic projection using $\kappa$-coordinates. Let us state the standard
transformation formulas for stereographic projection from \cite{Heckman}.

\smallskip

\begin{theorem}[Heckman-de Laat]
Stereographic projection $T^{*}\mathbb{R}^d \rightarrow T^{*}S^d$ is given by 
\begin{eqnarray}
X_k &=& \frac{u_k}{1-u_{d+1}}, \quad \quad \text{with} \, \, k=1,2,\cdots,d. \label{ster1}\\ 
X_k &=& v_k(1-u_{d+1}) + v_{d+1} u_k . \label{ster2}
\end{eqnarray}
and the inverse transformation formulae are given by
\begin{eqnarray}
u_k = \frac{2X_k}{X^2 +1}, \quad \quad \quad u_{d+1} = \frac{X^2-1}{X^2 +1} \label{inster1}\\
v_k = \frac{(X^2 + 1)y_k}{2} - <X,X> X_k, \quad \quad \quad v_{d+1} = <X,X>. \label{inster2}
\end{eqnarray}
The symplectic forms then satisfy 
\begin{equation}
{\sum_{k=0}}^{d} Y_k \wedge dX_k = {\sum_{k=0}}^{d} v_k \wedge du_k
\end{equation}
and hence the transformations are canonical.
\end{theorem}

\noindent
{\sl Proof}\,\, See \cite{Heckman}. $\Box$
 
\bigskip

A brief comment on the canonical transformations would be appropriate here. We consider a system described in terms of two different $\kappa$-coordinates, $(\psi,\phi)$ and $(\hx,\hp)$. Let $(\psi,\phi)$ be realized in terms of commutative variables $(x,p)$ and similarly, $(\hx,\hp)$  expressed in terms of $(y,q)$. Consider the realization of $\kappa$-space-time coordinates in terms of commuting variables (using notation introduced in the second section) are as given in eqns. \eqref{hatcor} and \eqref{hatmom}.

With the choice $a^{\mu} = (a, \vec{0})$, we write
\begin{eqnarray}
\psi^i &=& x^{i}+\alpha x^{i}(a p_0)+\beta(a x_0) p^{i}, \\
\phi^i &=& p^{i}+(\alpha +\beta)(a p_0)p^{i}. 
\end{eqnarray}
Recall that $\alpha - \gamma =1$ and $\beta$ can be any real number. We exploit this arbitrariness in the choice of $\beta$ to set it to zero. In other words, in the future we always assume $\beta = 0$, unless otherwise mentioned. The above expression can then be simplified to the form
\begin{eqnarray}
\psi^i &=& x^{i}+\alpha x^{i}(a p_0), \label{psia}\\
\phi^i &=& p^{i}+\alpha (a p_0)p^{i}.\label{phia}
\end{eqnarray}
 It is to be emphasized that the above realization obtained is exact and is valid to all orders in $a$. In the classical domain, we can think of the above set of equations as a "special" subset of canonical transformations provided, since in that case
\be
\{ \psi^i, \psi^j \}_{(x,p)} = 0, \quad  \{ \psi^i, \phi^j \}_{(x,p)} = \delta_{ij}, \quad \{ \phi^i, \phi^j \}_{(x,p)} = 0. \label{pbkappa}
\ee
The adjective ``special" is used above because the transformation have a dependence on the deformation parameter and $\{, \}$ refers to the Poisson bracket. An analogous treatment can be carried out for the case of $(\hx, \hp)$ also.

At classical level, any function $F(\psi, \phi)$ can be expressed as function in the $(x,p)$ phase space as $F(\psi, \phi) = f(x,p,a)$ where we will have a dependence on the deformation parameter `$a$'. Now, if we have canonical transformation connecting $(x,p)$ and $(y,q)$, then using the realization given by eqns. \eqref{psia} and \eqref{phia} we can establish a canonical transformation between $(\hX, \hY)$ and $(\hx, \hp)$. This in turn assures that the system described in terms of either $(\psi,\phi)$ or $(\eta, \zeta)$ refers to the same physical system. Another interesting point to be noted is that the canonical nature of transformations at the level of $\kappa$-spacetime can be understood by studying the coordinates in $\kappa$-spacetime as functions of ordinary phase space. Thus, the familiar Poisson bracket structure can be used to evaluate the coordinate transformation in $\kappa$-deformed phase space. These ideas are illustrated using the following diagram. 

\begin{figure}[!htbp]
\centering
\includegraphics[scale=.5]{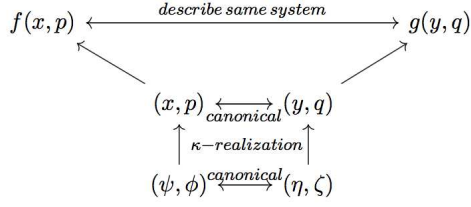}
\caption{Canonical transformations in $\kappa$-spacetime}\label{fig: canokappa}
\end{figure}

We start by considering a pair $(\phi, \psi) \in T^{*}(\kappa_M)^{d}$ satisfying relation \eqref{hatcor} and \eqref{hatmom}, where $\alpha,\beta,\gamma$ satisfy relations as in eqns.\eqref{abc}. 
\begin{eqnarray}
\psi_j &=& \frac{u_j}{1-u_{d+1}}, \quad \quad \text{with} \, \, j=1,2,\cdots,d. \label{kster1}\\ 
\phi_j &=& v_j (1-u_{d+1}) + v_{d+1} u_j . \label{kster2}
\end{eqnarray}
and the inverse transformation formulae are given by
\begin{eqnarray}
u_j &=& \frac{2\psi_j}{\psi^2 +1}, \quad \quad \quad u_{d+1} = \frac{\psi^2-1}{\psi^2 +1} \label{kinster1}\\
v_j &=& \frac{(\psi^2 + 1)y_j}{2} - <\psi,\phi> \psi_j, \quad \quad \quad v_{d+1} = <\psi,\phi>. \label{kinster2}
\end{eqnarray}
Few comments about the notations would be essential. 
\begin{enumerate}
\item It is to noted that the variables $(u,v)$ and $(x,p)$ are commutative, while $(\phi, \psi)$ are non-commutative variables.
\item Here we are defining a mapping from phase space $(u,v)$ of the $\ts$ to a phase space $(\hX,\hY)$. 
\item When we say $||\hX||^{2}$, we mean the norm defined in terms of the commuting variables appearing in the expression for $\hX$ given in eq. \eqref{kster1}, i.e., $||\hX||^{2} = \hX^{2} := ||x^{i}+\alpha x^{i}(a \cdot p)+\beta(a\cdot x)p^{i}||^{2}$ and similarly for $||\hY||$ (given by eq. \eqref{kster2}). 
\item  From now onwards, any calculus operation on $\hX$ actually means an operation on the commuting variables in terms of which it is expressed, i.e.,
\begin{eqnarray}
\frac{df(\hX)}{d\hX^{j}} := \frac{df(x,p)}{dx^{l}}\frac{dx^{l}}{d\hX^{j}}+\frac{df(x,p)}{dp^{l}}\frac{dp^{l}}{d\hX^{j}}
\end{eqnarray}
where $f$ is a function in $\kappa$-deformed phase space expressed in terms of commutative coordinates $(x,p)$.
\end{enumerate}

Coming back to our problem, consider now the geodesic flow on the manifold described by the pair $(u,v)$. We can define the Hamiltonian of the unit sphere 
\begin{equation}
F(u,v) = \frac{1}{2}|u|^2|v|^2 = \frac{1}{2}|v|^2 \label{ham1}
\end{equation}
The corresponding Hamilton's equations of motion would be given by
\begin{eqnarray}
u^{'} &=& F_v = v \label{hamf1}\\ 
v^{'} &=& F_u = - |v|^{2} u \label{hamf2}
\end{eqnarray}
where $F_u$, $F_v$ are the differentials of $F$ with respect to $u$ and $v$ respectively. These equations can be expressed as a second-order equation of the form,
\be
u^{''} + \abs{u'}^{2} u = 0.
\ee

We will now proceed to initiate the mapping between $\ts$ to $\tm$. For this, we first restrict ourselves to geodesic flow with $|v|=1$, in other words we focus on the hypersurface $F=\frac{1}{2}$(see eqn.\eqref{ham1}). The Hamiltonian, in terms of non-commutative coordinates, will then be
\be 
F(\psi,\phi) = F(u,v) = \frac{(\psi^2 +1)^2 \phi^2}{8}. \label{ham2}
\ee
Note that the re-expressed Hamiltonian has a dependence on the deformation parameter through eqns.\eqref{kster1},\eqref{kster2}. Recall that our Hamiltonian equations of motion would then be given by
\begin{eqnarray}
\hX^{'}=F_{\hY}, \quad \hY^{'} = -F_{\hX}, \label{eomham}
\end{eqnarray}
and they correspond to $F=\frac{1}{2}$. Since only the gradient of $F$ is relevant for the above differential equations, any function of $F$, $g(F)$ will also be sufficient, provided $g^{'}(\frac{1}{2})=1$ because the gradient of the two functions ($F$ and $g(F)$) agrees on hypersurface $F=\frac{1}{2}$. 

We will now modify the the above Hamiltonian $F(\hX,\hY)$ in such a manner that we will end up with a Hamiltonian having the same set of equation of motion as in eqn \eqref{eomham}. Having set the stage, we can perform change of parametrisation to arrive at the Hamiltonian of the Kepler problem, which is what we are looking for. For this, consider a modified Hamiltonian expressed in terms of $\hX$ and $\hY$, given by
\begin{eqnarray}
G(\hX,\hY) = \sqrt{2F(\hX,\hY)} - 1 = \frac{({\hX}^{2}+1)\hY}{2} - 1
\end{eqnarray}
The equation of motion on $F=\frac{1}{2}$ goes to 
\begin{eqnarray}
\hX^{'}=G_{\hY}|_{G=0}, \quad \hY^{'} = -G_{\hX}|_{G=0}.
\end{eqnarray}
A change of evolution parameter from $s$ to a deformation dependent parameter, $t_a$ ($a$ indicates the dependence on deformation) is provided by 
\begin{equation}
t_a = \int |\hY| \, ds
\end{equation}
\underline{Note:} $|\hY| = \sqrt{\hY^{2}}$.

 A comment about the differentiation w.r.t. $t_a$ would be appropriate. The variable $t_a$ acts like the usual parameter of evolution except that it have an additional dependence on the deformation parameter. We define the evolution of a function in phase space, w.r.t. $t_a$ by
\be 
\frac{d \mathcal{A}}{d t_a} = - [\mathcal{A}, G].
\ee
where $\mathcal{A}$ is an arbitrary function in the concerned phase space. This is to be contrasted with the approach taken in \cite{luk1} where differentiation w.r.t. time coordinate of \kdst \, is shown to be a finite difference operation.

\bigskip
If we now denote the differentiation w.r.t. $t_a$ by an overdot, then the equation of motion associated with $G$ can be re-expressed as
\begin{eqnarray}
\dot{\hX}=\frac{1}{|\hY|}G_{\hY}, \quad \dot{\hY} = -\frac{1}{|\hY|}G_{\hX}\quad \quad \text{on} \, \, G=0.
\end{eqnarray}
Redefine,
\begin{eqnarray}
\frac{1}{|\hY|}G_{\hY} = H_{\hY}, \quad \quad \frac{1}{|\hY|}G_{\hX} = H_{\hX} \quad \quad \text{on} \, \, H= -\frac{1}{2}.
\end{eqnarray}
where
\begin{equation}
H = \frac{1}{|\hY|} G - \frac{1}{2} = \frac{1}{2} |\hX|^{2} - \frac{1}{|\phi|}. \label{phiham}
\end{equation}. 

This redefinition allows us to express the above equation of motion as arising from a Hamiltonian function. Further, we could make an identification of this Hamiltonian with the Hamitonian for a Kepler problem. 
If we let $\hX = \hp$ and $\hY = \hx$ in eq.\eqref{phiham}, then $(\hX,\hY) \mapsto (\hp,\hx)$ and the corresponding Hamiltonian coincide with the Hamiltonian for the Kepler system,
\begin{equation}
H(\hx,\hp) = \frac{\hp^{2}}{2} - \frac{1}{|\hx|}
\end{equation}

 We have thus achieved a mapping from geodesic flow on a $d$-dimensional sphere to a hypersurface $H=-\frac{1}{2}$ on a Kepler Hamiltonian defined on $d$-dimensional $\kappa$-spacetime.
 \section{Ligon-Schaaf regularization}
Ligon-Schaaf (LS) map provides a nice way to regularize the negative energy Keplerian orbits all at once. LS map is a canonical transformation that connects bounded Kepler orbits as a whole to the geodesic orbits of the sphere. It is to be noted that unlike Moser method, there is no reparametrisation of evolution of parameter. In addition, the constants of motion of the sphere is transformed to the conserved quantities of the Kepler problem. For more details see \cite{LS,cush1,cush2}.

\subsection{Review of regularization of Kepler problem in commutative spacetime}

In order to carry out the regularization of Kepler problem, we first need to set up the dynamics for motion on a sphere. Once we have the Hamilton's equations for the motion of particle on a sphere, we will be able to use the LS-map to connect the dynamics on sphere with the dynamics of Kepler problem. It is to be emphasized that we will map a sphere in ordinary commutative spacetime to the deformed Kepler problem. Hence, all the coordinates referring to the sphere would be commutative where as all those referring to Kepler problem belongs to $\kappa$-deformed phase space expressed in terms of commutative variables.

We will first briefly look at the problem of particle on a sphere. Let us denote the phase space of 3-sphere by $TS^{3}$ with coordinates $(x,y)$. Also, $T^{+}S^{3} := \{ (x,y) \in TS^{3} \mid y\neq 0\}$., On $(T^{+}S^{3}\subset T\rn^{4}, \tilde{\omega}_{4})$ consider the Delaunay Hamiltonian:
\be 
\tilde{H} = - \frac{\mu^{2}}{2<y,y>}
\ee
where $\mu$ will be later identified with our modified mass (see eqn. \eqref{mass}). We will have the constraint relations given by
\be 
c_1: (x,y) \mapsto \frac{1}{2}(<x,x>-1), \quad \quad c_2: (x,y) \mapsto <x,y>.
\ee
Since the matrix of Poisson bracket $\{c_i, c_j \}$ is invertible, we can have the inverse matrix as 
\be 
C_{ij} = \frac{1}{<x,x>}\left( \begin{array}{cc}
0 & -1 \\
1 & 0
\end{array}\right)
\ee
Using modified Dirac bracket procedure, we can write the modified Hamiltonian function $H^{*}$ as
\be 
H^{*} = \tilde{H} - \sum_{i,j=1}^{2} (\{\tilde{H},c_i \}+ \tilde{H}_i) C_{ij}c_j
\ee
where we choose, for convenience, 
\be 
\tilde{H}_1 = <x,y>(\frac{\mu^{2}}{<y,y>^{2}}-\frac{1}{2}<x,x>), \quad \tilde{H}_2 = \frac{\mu^{2}}{<y,y>} (<x,x>-1).
\ee
Then,
\be 
H^{*} = -\frac{\mu^{2}}{2<y,y>} - <x,y>^{2} + \frac{1}{2} \frac{\mu^{2}}{<y,y>}(<x,x,>-1).
\ee
So Hamilton's equations of motion are given by
\begin{eqnarray}
\dot{x} = \frac{\partial H^{*}}{\partial y} =  \frac{\mu^{2}}{<y,y>^{2}}y, \\
\dot{y} = \frac{\partial H^{*}}{\partial x} = - \frac{\mu^{2}}{<y,y>} x.
\end{eqnarray}

Before turning to deformed case, we will have an overview on the LS-map in the usual ($a=0$) case (see \cite{LS}).

We set the LS regularization map as $\Phi_{LS}(q,p)= (x,y)$, with
\be 
(x,y) = (A\sin\theta + B\cos \theta, -v A\cos \theta + vB\cos \theta),
\ee
where
\begin{eqnarray}
A &=& (\tilde{A}, A_4) = (\frac{q}{\vert q \vert}- \frac{1}{\mu}<q,p>p, \frac{1}{v}<q,p>), \\
B &=& (\tilde{B}, B_4) = (\frac{1}{v}\vert q \vert p, \frac{1}{\mu}<p,p>\vert q \vert - 1),
\end{eqnarray}
\be 
\text{with} \quad v = \sqrt{\frac{\mu}{-2H}}.
\ee
{\bf Claim:} LS map intertwines the momentum map of Kepler vector field and Delaunay vector field. 
\bigskip

{\bf Proof:} We will now show that the 
\be 
(1). \, \tilde{x} \times \tilde{y} = q \times p \quad \text{and} \quad (2). \, x_4 \tilde{y} - y_4 \tilde{x} = v \left[ \frac{q}{\vert q \vert} - \frac{1}{\mu}p\times (q \times p)\right].
\ee
\begin{eqnarray}
1). \,  \tilde{x} \times \tilde{y} &=& (\tilde{A}\sin \theta + \tilde{B} \cos \theta) \times v(-\tilde{A} \cos \theta + \tilde{B} \sin \theta), \\
&=& v (\ta \times \tb \sin^{2}\theta + \ta \times \tb \cos^{2} \theta ) + 0, \\
&=& v (\frac{q}{\vert q \vert} \times \frac{1}{v}\vert q \vert p), \\
&=& q \times p.
\end{eqnarray}
\begin{eqnarray*}
2). \, x_4 \tilde{y} - y_4 \tilde{x} &=& \left(\fa \sin \theta + \fb \cos \theta)(-v \ta \cos \theta + v \tb \sin \theta)\right.\\ && \left. - (-v \fa \sin \theta + v \fb \sin \theta) (\ta \sin \theta + \tb \cos \theta)\right), \\
&=& v \fa \tb \sin^{2} \theta - v \fb \ta \cos^{2} \theta + v \fa \tb \cos^{2} \theta - v \fb \ta \sin^{2} \theta, \\
&=& v \fa \tb - v \fb \ta, \\
&=& v \left( \frac{1}{v^{2}} <q,p> \vert q \vert p - (\frac{1}{\mu}<p,p>\vert q \vert - 1)(\vert q \vert - \frac{1}{\mu}<q,p>p)\right)
\end{eqnarray*}
We will now use the expression, $2H = \frac{<p,p>}{\mu} - \frac{2}{\vert q \vert }$, to obtain
\begin{eqnarray*}
\, x_4 \tilde{y} - y_4 \tilde{x} &=& v \left( (\frac{- \vert q \vert <p,p>}{\mu^{2}}+2)<q,p> p - (\frac{\vert q \vert <p,p>}{\mu}-1)\frac{q}{\vert q \vert} 
\right.\\ && \left. +  (\frac{\vert q \vert <p,p>}{\mu^{2}} -1) <q,p> p \right), \\
&=& v <q,p> p - v ( \frac{<p,p>}{\mu} - \frac{1}{\vert q \vert }) q, \\
&=& v (\frac{q}{\vert q \vert} - \frac{1}{\mu}p\times (q \times p)).
\end{eqnarray*}

\subsection{Review of Kepler problem in kappa spacetime}
  Now let us consider the Kepler potential $V(\psi^i ) = -\frac{C}{\hat{r}}$ where $\hat{r}=\sqrt{\psi^i \psi_i}$, where $\psi$ is given by eqn. \eqref{hatcor}. Expressing the potential in terms of ordinary spacetime variables, using eqns. \eqref{hatcor} and \eqref{hatmom}, we obtain (see \cite{zns1, zns2} for details).
  \begin{equation}
   V(r) = -\frac{C}{r} (1+a\alpha p_0) 
  \end{equation}
 Note that the expression is exact in the sense that it is valid to all orders in $a$. 
 
 The corresponding Hamiltonian can be written as
  \begin{eqnarray}
  H&=&\frac{p_{i}^2}{2\tilde{\mu}}-\frac{C}{r}(1+a\alpha  p_0) \\
    &=&\frac{p_{i}^2}{2\tilde{\mu}}-\frac{\tilde{C}}{r} \label{hamk}
  \end{eqnarray}
  where $\tilde{C}= C(1+a\alpha  p_0)$ and the deformed mass is given by.
\be   
  \tilde{\mu}=m/(1+a\alpha m) \label{mass}
  \ee
  Thus the \kd \, Hamiltonian describing the Kepler problem given in eqn.\eqref{hamk} is valid to all orders in the deformation parameter $a$.
  
  \bigskip
  
   We see that the integrals of motion are the modified angular momentum and Laplace-Runge-Lenz vector given by 
  \begin{eqnarray}
   \hat{L} &=&  \vec{r}\times \vec{p}, \\ 
     \hat{A} &=&  \frac{1}{\tilde{\mu}} \vec{p}\times \hat{L} -  \tilde{C}\frac{\vec{r}}{r}.
     \end{eqnarray}
Notice that the deformation in $\hat{L}$ arises from the deformation of mass term present in the definition of momentum $\vec{p}=\tilde{\mu} \vec{v}$. Similarly, for Runge-Lenz vector, the deformation arises from the factors $\tilde{\mu}$ and $\tilde{C}$.  
     
\bigskip
     
We will denote the negative energy surface by $\Sigma_{-}$ and will have the following definition.

{\bf Definition:} 
\be 
\Sigma_{-} := \{(q,p) \mid H(q,p) < 0 \}
\ee

For definiteness we will denote the symplectic form of Kepler problem by $\omega_{3} = \sum_{i=1}^{3} dq_i \wedge dp_i $ and let $\tilde{\omega}_{3} = \omega_{3}\vert_{ \Sigma_{-}}$. Note that the phase space we are working is the usual commutative phase space, with coordinates $(q,p)$ and all our functions in non-commutative variables would be written as functions of the commutative variables with a dependence on the deformation parameter. So for all computational purposes, we can view ourselves working on the $(q,p)$ phase space.

On $(\Sigma_{-}, \tilde{\omega_{3}})$, $\vec{L}$ \, \, and $\vec{B}$ satisfy the Poisson bracket relations
\be 
\{L_i, L_j \} = \epsilon_{ijk} L_k, \quad \{L_i, B_j \} = \epsilon_{ijk} B_k \text{and} \quad \{B_i, B_j \} = \epsilon_{ijk} L_k
\ee
where $\vec{B} = \frac{\mu}{\sqrt{-2H}} \vec{A}$.

{\bf Claim:} The above Lie algebra is isomorphic to $so(4)$ algebra. \\

{\bf Proof:} Define: $U_i := \frac{1}{2} (L_i + B_i), V_i := \frac{1}{2} (L_i - B_i)$. Then the poisson bracket relations become
\be 
\{U_i, U_j \} = \epsilon_{ijk} U_k, \quad \{V_i, V_j \} = \epsilon_{ijk} V_k \, \, \text{and} \, \, \{U_i, V_j \} = 0.
\ee
These relations define the $so(3) \times so(3)$ algebra which is isomorphic to $so(4)$ algebra. This implies that the deformed-Kepler system possess same symmetry algebra as in the undeformed situation.

\subsection{LS regularization on $\kappa$-deformed space}

Having studied the regularization scheme in commutative situation we will now demonstrate the mapping in the deformed case. For this, we will be following the notation set up in section 3. 
\be
(X,Y) = (A\sin\theta + B\cos \theta, -v A\cos \theta + vB\cos \theta), 
\ee
where
\begin{eqnarray}
A &=& (\tilde{A}, A_4) = (\frac{\psi}{\vert \psi \vert}- \frac{1}{\mu}<\psi ,\phi>\phi, \frac{1}{v}<\psi ,\phi>), \label{ALS}\\
B &=& (\tilde{B}, B_4) = (\frac{1}{v}\vert \psi  \vert \phi, \frac{1}{\mu}<\phi, \phi>\vert \psi  \vert - 1), \label{BLS}
\end{eqnarray}
with $v = \sqrt{\frac{\tilde{\mu}}{-2H}}$ and $2H = \frac{\phi^{2}}{\tilde{\mu}} - \frac{2}{\vert \psi \vert}$. It should be emphasized that the LS map have a dependence on the deformation parameter, $a$ through the realization of $\phi$ and $\psi$ in terms of the commuting variables. (see eqns. \eqref{hatcor} and \eqref{hatmom}).

{\bf Claim:} LS map intertwines the momentum map of Kepler vector field and Delaunay vector field. 
\bigskip

{\bf Proof:} We will now show that the 
\be 
(1). \, \tilde{x} \times \tilde{y} = \hX \times \hY \quad \text{and} \quad (2). \, x_4 \tilde{y} - y_4 \tilde{x} = v \left[ \frac{\hX}{\vert \hX \vert} - \frac{1}{\tilde{\mu}}\hY\times (\hX \times \hY)\right].
\ee
\begin{eqnarray}
1). \,  \tilde{x} \times \tilde{y} &=& (\tilde{A}\sin \theta + \tilde{B} \cos \theta) \times v(-\tilde{A} \cos \theta + \tilde{B} \sin \theta), \\
&=& v (\ta \times \tb \sin^{2}\theta + \ta \times \tb \cos^{2} \theta ) + 0, \\
&=& v (\frac{\hX}{\vert \hX \vert} \times \frac{1}{v}\vert \hX \vert \hY), \\
&=& \hX \times \hY.
\end{eqnarray}
\begin{eqnarray}
2). \, x_4 \tilde{y} - y_4 \tilde{x} &=& (\fa \sin \theta + \fb \cos \theta)(-v \ta \cos \theta + v \tb \sin \theta) - (-v \fa \sin \theta + v \fb \sin \theta) (\ta \sin \theta + \tb \cos \theta), \nonumber \\
&=& v \fa \tb \sin^{2} \theta - v \fb \ta \cos^{2} \theta + v \fa \tb \cos^{2} \theta - v \fb \ta \sin^{2} \theta, \nonumber \\
&=& v \fa \tb - v \fb \ta, \nonumber \\
&=& v \left( \frac{1}{v^{2}} <\hX,\hY> \vert \hX \vert \hY - (\frac{1}{\tilde{\mu}}<\hY,\hY>\vert \hX \vert - 1)(\vert \hX \vert - \frac{1}{\tilde{\mu}}<\hX,\hY>\hY)\right)
\end{eqnarray}
We will now use the expression, $2H = \frac{<\hY,\hY>}{\mu} - \frac{2}{\vert \hX \vert }$, to obtain
\begin{eqnarray}
\, x_4 \tilde{y} - y_4 \tilde{x} &=& v \left( (\frac{- \vert \hX \vert <\hY,\hY>}{\tilde{\mu}^{2}}+2)<\hX,\hY> \hY - (\frac{\vert \hX \vert <\hY,\hY>}{\mu}-1)\frac{\hX}{\vert \hX \vert}\right) \nonumber\\ 
&& \left. +  (\frac{\vert \hX \vert <\hY,\hY>}{\tilde{\mu}^{2}} -1) <\hX,\hY> \hY \right) \nonumber \\
&=& v <\hX,\hY> \hY - v ( \frac{<\hY,\hY>}{\tilde{\mu}} - \frac{1}{\vert \hX \vert }) \hX, \nonumber\\
&=& v (\frac{\hX}{\vert \hX \vert} - \frac{1}{\tilde{\mu}}\hY \times (\hX \times \hY)).
\end{eqnarray}
Indeed, LS map intertwines the vector field of $\kappa$-deformed Kepler and Delaunay as in the commutative situation. Note that $\phi$ and $\psi$ appearing in the above eqn. are \kd \, phase space coordinates given in eqn. \eqref{hatcor} and \eqref{hatmom}. These are valid to all orders in the deformation parameter $a$.

Geometrically, we consider $T^{+}S^{3}$ as a regularized model for the negative energy subset $\Sigma_{-}$ of the phase space of the Kepler problem. On $T^{+}S^{3}$  the SO(4) symmetry is globally defined and its orbits
\[
\{(x,y) \in T^{+}S^{3} \mid <y,y> = - \frac{\tilde{\mu}}{2H}  \}
\]
are the regularization of negative energy surface of Kepler problem. Further note that the mass appearing in the above equation for the phase space of a sphere is numerically equal to the deformed mass. Expressed differently, the mass of particle in motion on a sphere have to be identified with the deformed mass of the equivalent Kepler problem. This identification is necessary for the LS-regularization to be consistent.

The above calculation is exactly analogous to the commutative ($a=0$) situation and hence, have been extended to $\kappa$ case($a\neq 0$). 

\section{Discussion and Conclusion}
In this paper, we have studied the regularization of $\kappa$-deformed Kepler problem. We used an extension of Moser method to regularize Kepler problem in $\kappa$-spacetime. The mapping of geodesic motion on a sphere to the Kepler problem in deformed-spacetime have dependence on the deformation parameter `$a$'. It should be reminded that this dependence arise not only from realization of the non-commutative coordinates in terms of the commutative coordinates, but also from the possible dependence of the parameter of evolution on the deformation parameter (see eqns. \eqref{ALS},\eqref{BLS}). We have studied the generalization of the Ligon-Schaaf map to the $\kappa$-deformed case and it is established that the procedure of regularization can be carried out in a straightforward manner, as in the commutative situation. The key idea in our regularization is that we consider coordinates of $\kappa$-deformed phase space in terms of functions of usual variables in Minkowski phase space. Note that we use the realization of the $\kappa$-deformed phase space variables in terms of the commutative ones and employ Poisson bracket relation (not quantum commutators) in this paper. The incompleteness in the flow of Hamiltonian vector field is rectified by embedding it into a complete flow. We have extended the regularization procedure in commutative spacetime to the case of deformed situation. 

Our starting point of analysis is the deformed Kepler problem, defined by the Hamiltonian given in eqn. \eqref{hamk} for the 
generalisation of L-S method to \kdst . For extending the Moser method, eqns. \eqref{psia}, \eqref{phia} and \eqref{kster1}, 
\eqref{kster2} are the starting points. Note that in the limit $a \rightarrow 0$, these equations reduce to the correct commutative 
limit. It is possible to have many other generalisations of Kepler problem to \kdst\, which, in the commutative limit, will reproduce 
the expected result. Thus whether the model studied here is related to \kd \, Riemannian geometry is to be shown. This study is in
progress and will be reported separately.We have shown here that the extension of regularization methods of Kepler problem to the 
\kdst \, is possible by explicitly constructing the Moser and L-S regularizations for the \kd \, Kepler problem.

\bigskip

\noindent{\bf Acknowledgments.} We thank the anonymous referee for useful suggestions. PG would like to thank the Tudor Ratiu very much for discussion and ongoing collaboration on geometry of Kepler equation. ZNS would like to acknowledge the support for this work from CSIR, India through the SRF scheme. EH thanks SERB, Govt. of India, for support through EMR/2015/000622.

\end{document}